\documentclass[conference]{IEEEtran}
\usepackage{latexsym}
\usepackage{graphicx}
\usepackage{amsfonts,amssymb,amsmath}
\usepackage{algorithm,algorithmic}
\usepackage{varwidth}% http://ctan.org/pkg/varwidth
\usepackage{hyperref}

\usepackage[T1]{fontenc}
\usepackage{cite}
\usepackage{comment}
\usepackage{amsthm}
\usepackage{overpic}
\usepackage{steinmetz}
\usepackage{array}
\usepackage{url}
\usepackage{xcolor}
\IEEEoverridecommandlockouts

\theoremstyle{plain}

\newcommand{\vect}[1]{\mathbf{#1}}

\def\diag{\mathrm{diag}}

\def\Htran{\mbox{\tiny $\mathrm{H}$}}
\def\Ttran{\mbox{\tiny $\mathrm{T}$}}
\def\CN{\mathcal{N}_{\mathbb{C}}} %Complex Gaussian
\def\H{\textrm{H}}
\def\T{\textrm{T}}
\def\V{\textrm{V}}

\begin{document}
\bstctlcite{IEEEexample:BSTcontrol} % to shrink citations

\title{Efficient Multi-Source Localization in Near-Field Using only Angular Domain MUSIC}

\author{Mehdi Haghshenas, Aamir Mahmood, Mikael Gidlund  \\
\IEEEauthorblockA{\textit{Department of Computer and Electrical Engineering, Mid Sweden University, 851 70 Sundsvall, Sweden} \\ Email: \{mehdi.haghshenas, aamir.mahmood, mikael.gidlund\}@miun.se}
%\IEEEmembership{Graduate Student Member, IEEE}, TBA\\
%\thanks{}%
%\thanks{}% 
\vspace{-0.7cm}
}

\maketitle

\begin{abstract}
The localization of multiple signal sources using sensor arrays has been a long-standing research challenge. While numerous solutions have been developed, signal space methods like MUSIC and ESPRIT have gained widespread popularity. As sensor arrays grow in size, sources are frequently located in the near-field region. The standard MUSIC algorithm can be adapted to locate these sources by performing a 3D search over both the distance and the angles of arrival (AOA), including azimuth and elevation, though this comes with significant computational complexity. To address this, a modified version of MUSIC has been developed to decouple the AoA and distance, enabling sequential estimation of these parameters and reducing computational demands. However, this approach suffers from reduced accuracy. To maintain the accuracy of MUSIC while minimizing complexity, this paper proposes a novel method that exploits angular variation across the array aperture, eliminating the need for a grid search over distance. The proposed method divides the large aperture into smaller sections, with each focusing on estimating the angles of arrival. These angles are then triangulated to localize the sources in the near-field of the large aperture. 
Numerical simulations show that this approach not only surpasses the Modified MUSIC algorithm in terms of mean absolute error but also achieves accuracy comparable to standard MUSIC, all while greatly reducing computational complexity—$370$ times in our simulation scenario.
\end{abstract}
\begin{IEEEkeywords}
Near-Field, Source Localization, MUSIC Algorithm
\end{IEEEkeywords}

\maketitle

\section{Introduction}
The detection and localization of multiple sources using sensor arrays has been a longstanding area of research~\cite{1996twodecadesignal}. Numerous solutions have been proposed, focusing on the localization of sources within either the far-field or near-field regions of the sensor array~\cite{1969Capon,1986MUSIC,1989ESPRIT,1990MLE,1991NFMUSIC,2002MLENF,2005GeneralizedESPRIT,2005LinPredictionNF,2006RMQMUSIC,2007ModifiedMUSIC,2009NFFFMUSIC,2012FFNFMUSIC,2018NFMUSIC}. These approaches include non-parametric methods such as the Capon method~\cite{1969Capon}, parametric Maximum Likelihood~\cite{1990MLE}, compressed sensing techniques~\cite{2016CSReview}, and subspace-based methods such as MUltiple SIgnal Classification (MUSIC) and Estimation of Signal Parameters via Rotational Invariance Techniques (ESPRIT)~\cite{1986MUSIC,1989ESPRIT}. Among these, the MUSIC algorithm has become a popular and widely used solution for localizing multiple concurrent sources~\cite{2024ramezanibookchapterlocalization}. 

The MUSIC algorithm conducts a high-resolution search across the angular plane to estimate the angle of arrival (AoA) for multiple far-field sources~\cite{1986MUSIC}. However, with the increasing number of array elements, signal sources often fall within the near-field region, where the wavefront can no longer be approximated as planar. While MUSIC can be extended to perform a search over both angle and distance, this becomes computationally impractical. As a result, several methods have been developed to localize sources in the near-field~\cite{2002MLENF,2009NFFFMUSIC,2012FFNFMUSIC,2018NFMUSIC}. For example, in~\cite{2012FFNFMUSIC}, He \textit{et al.} proposed a Modified MUSIC algorithm that leverages array symmetry to decouple angle and distance estimation. This method utilizes the anti-diagonal elements of the received correlation matrix to form a semi-MUSIC spectrum, estimating AoAs by identifying peaks in the high-resolution angular search. In the subsequent step, for each estimated angle, the distance is determined by conducting an exhaustive search over the distance using the standard MUSIC spectrum. Although this approach reduces computational complexity, it does so at the expense of localization accuracy.

This paper introduces a novel, computationally efficient method based on the MUSIC algorithm for localizing multiple sources in the near-field region of a sensor array. In large arrays, when a source is positioned in the near-field, the angle of arrival varies across different subsets of sensors. The proposed approach exploits this angular variation to precisely localize multiple sources without the need for distance estimation. By reducing the search space from the angle-distance domains to the angle domain alone, the method significantly lowers computational complexity compared to Modified MUSIC, while achieving better performance. Moreover, the proposed solution achieves accuracy comparable to standard MUSIC, highlighting both its efficiency and effectiveness.

The structure of this paper is as follows: Section~\ref{sec:sysmod} introduces the system model and array response for a uniform linear array (ULA) and a uniform planar array (UPA) when the source is located in either the far-field or near-field. Section~\ref{sec:MUSIC} explains the theory behind the MUSIC algorithm, which will later be applied in our proposed method. Section~\ref{sec:proposed_localization_method} outlines the proposed localization approach, and Section~\ref{sec:numericalRes} presents and discusses the performance and efficiency of the algorithm. Finally, Section~\ref{sec:conclusion} provides the conclusion of the paper.

\subsection{Notations}
Vectors and matrices are
respectively denoted by boldface lowercase and boldface uppercase
letters. $(\cdot)^{\Htran}$ and $(\cdot)^{\Ttran}$ denote Hermitian transpose and transpose, respectively. $\diag(a_1,\ldots,a_N)$ is a diagonal matrix having $a_1,\ldots,a_N$ as its diagonal elements. The symbol $\mathbb{C}$ denotes the complex set and the distribution of a circularly symmetric complex Gaussian random
variable with variance $\sigma^2$ is denoted
by $\sim \CN(0, \sigma^2)$, where $\sim$ stands for “distributed as”. The symbol $\in$ indicates set membership.

%%%%%%%%%%%%%%%%%%%%%%%%%%%%%%%%%%%%%%%%%%%%%%%%%%%%%%%%%%%%%%%%%%%%%%%%%%%%%%%%%%%%%%%%%%%%%%%%%%%%%%%%%%%%%%%%%%%%
\section{System Model}
\label{sec:sysmod}

We consider an array of $M$ sensors
%arranged in a ULA configuration, with an inter-element spacing of $d$, 
receiving signals from $K$ sources. The received signal at the sensor array, denoted as $\vect{y} = [y_1, y_2, \dots, y_M]^\T \in \mathbb{C}^M$, can be expressed as:
\begin{equation}
\label{eq:rx_signal}
    \vect{y} = \sum_k^K\vect{h}_kx_k + \vect{n},
\end{equation}
where $\vect{h}_k \in \mathbb{C}^M$ represents the channel between the $k$-th source and the sensor array, and $x_k \sim \CN(0,\rho_k)$ is the signal from the corresponding source. Moreover, $\vect{n} \sim \CN(\vect{0},\sigma^2\vect{I}_M)$ represents the receiver's additive white Gaussian noise.

The channel for the $k$-th incoming signal can be described as:
%We assume that the system is operating in mmWave band where the channel can be approximated as LOS dominant. In this specific case, the channel for each signal source can be approximated as single path (LOS) and can be expressed as:
\begin{equation}
\label{eq:channel}
    \vect{h}_k = \sqrt{\beta_k} e^{-j\frac{2\pi}{\lambda}r_{1k}} \vect{a}(\boldsymbol{\psi}_k),  
\end{equation}
where $\lambda$ is the wavelength, $\beta_k \in \mathbb{R}$ is the channel gain, and $r_{1k}$ denotes the distance between the $k$-th source and the reference antenna element. The vector $\vect{a}(\boldsymbol{\psi}) \in \mathbb{C}^M$ is the array response, which contains the phase shifts relative to the reference element and is defined as:
\begin{equation}
\label{eq:general_array_response}
    \vect{a}(\boldsymbol{\psi}_k) = [1,\, e^{-j\frac{2\pi}{\lambda}(r_{2k} - r_{1k})}, \ldots, \, e^{-j \frac{2\pi}{\lambda}(r_{Mk} - r_{1k}) }]^\T.
\end{equation}
Here, the argument $\boldsymbol{\psi}_k$ is a function of $(\varphi_k,r_{1k})$ in the near-field of a ULA, and $(\varphi_k,\theta_k,r_{1k})$ in the near-field of a UPA, where $\varphi$ and $\theta$ represent the azimuth and elevation angles, respectively. In contrast, in the far-field, it depends solely on the angles and not the distance. 

Substituting \eqref{eq:channel} into \eqref{eq:rx_signal}, the received signal can be rewritten as: 
\begin{equation}
    \vect{y} = \vect{A}\vect{B}\vect{x} + \vect{n},
\end{equation}
where $\vect{x} = [x_1,x_2,\ldots,x_K]^\T$ contains the signals from the $K$ sources. The matrices $\vect{B} \in \mathbb{C}^{K \times K}$ and $\vect{A} \in \mathbb{C}^{M \times K}$ contains the channel's complex coefficients and the array responses, respectively, and are defined as:
\begin{equation}
    \vect{B} = \diag \left([\sqrt{\beta_1} e^{-j\frac{2\pi}{\lambda}r_{11}},\ldots, \sqrt{\beta_K} e^{-j\frac{2\pi}{\lambda}r_{1K}}]\right),
\end{equation} 
\begin{equation}
    \vect{A} = \left[\vect{a}(\boldsymbol{\psi}_1)^\T, \vect{a}(\boldsymbol{\psi}_2)^\T, \ldots, \vect{a}(\boldsymbol{\psi}_K)^\T \right].
\end{equation}
\subsection{Far-Field v.s. Near-Field}
\begin{figure}
    \centering
    \begin{overpic}[width=.7\columnwidth]{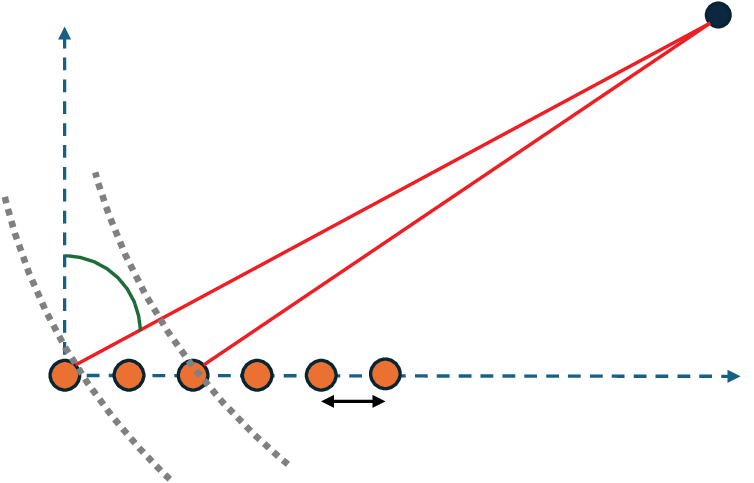}
        \put (9,32) {$\varphi_k$ }
        \put (60,60) {Transmitter $k$}
        \put (22,8) {$m$}
        \put(95,16){$Y$}
        \put(3,57){$X$}
        \put(60,48){$r_{1k}$}
        \put(69,40){$r_{mk}$}
        \put(45,6){$d$}
    \end{overpic}
    \caption{An uniform linear array with $M$ antennas and inter-element spacing of $d$ is spanned on Y-axis. The wave is propagating spherically from the transmitter toward the receiver.}
    \label{fig:sphericalwavefront}
\end{figure}
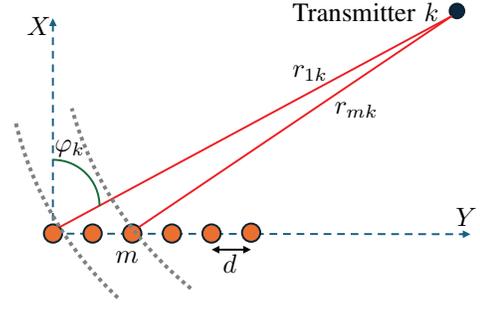

Figure~\ref{fig:sphericalwavefront} illustrates a wave propagating spherically from an arbitrary source $k$ to the ULA receiver. Using the law of cosines, the distance from the source to the $m$-th element in the array can be expressed as follows~\cite{1991NFMUSIC}:
\begin{equation}
\label{eq:distance_ULA}
    \begin{aligned}
        & r_{mk}   = r_{1k} \sqrt{1 + \frac{(m-1)^2d^2}{r_{1k}^2} - \frac{2(m-1)d\sin(\varphi_k)}{r_{1k}}} \\ 
        & \overset{(a)}{\approx} r_{1k} \!\left(\! 1 + \frac{(m-1)^2d^2}{2r_{1k}^2} - \frac{(m-1)d \sin(\varphi_k)}{r_{1k}} - \frac{(m-1)^4d^4}{8r_{1k}^4} \right. \\
        & \quad \left.- \frac{(m-1)^2d^2 \sin^2(\varphi_k)}{2r_{1k}^2} + \frac{(m-1)^3d^3\sin(\varphi_k)}{2r_{1k}^3}\right) \\ 
        & \overset{(b)}{\approx} r_{1k} \left( 1 + \frac{(m-1)^2d^2}{2r_{1k}^2}\cos^2(\varphi_k) - \frac{(m-1)d \sin(\varphi_k)}{r_{1k}}\right)
    \end{aligned}
\end{equation}
where $d$ denotes inter-element spacing, and $\varphi_k$ is the angle of arrival for source $k$, measured with respect to the $X$-axis. In step (a), a second-order Taylor approximation, $\sqrt{1+x} \approx 1 + \frac{x}{2} - \frac{x^2}{8}$ for $|x| \ll 1$ is applied, and (b) is derived under the assumptions $d^3 \ll r_{1k}^3$ and $d^4 \ll r_{1k}^4$. The final expression in~\eqref{eq:distance_ULA} is then used as input in~\eqref{eq:general_array_response} to compute the array response for a ULA in the near-field.

When the source is in the far-field of the sensor array, i.e., $(m-1)d \ll r_{1k}$, the spherical wavefront can be approximated as a planar wavefront. Consequently, \eqref{eq:distance_ULA} simplifies to:
\begin{equation}
    r_{mk} = r_{1k} - (m-1) d \sin(\varphi_k) \rightarrow r_{1k} - r_{mk} = (m-1)d\sin(\varphi_k).
\end{equation}
As a result, the array response in~\eqref{eq:general_array_response} can be simplified for sources in the far-field region. In this case, the array response depends only on the angle and is given by:
\begin{equation}
\label{eq:ULAarrayreponse}
    \vect{a}(\varphi) = [1, e^{j\frac{2\pi}{\lambda}d\sin(\varphi)},\ldots, e^{j\frac{2\pi}{\lambda}(M-1)d\sin(\varphi)}]^\T.
\end{equation}

\subsection{Extension to uniform planar array}
The array response in \eqref{eq:general_array_response} applies to any sensor array configuration. When the array is structured as a ULA, the distance $r_{mk}$ is given by \eqref{eq:distance_ULA}, resulting in the far-field array response in  \eqref{eq:ULAarrayreponse}.  However, for a UPA, \eqref{eq:distance_ULA} and \eqref{eq:ULAarrayreponse} must be replaced with the following equations~\cite{2024Haghshenas,2024ramezaniefficientmodifiedmusicalgorithm}:
\begin{equation}
\label{eq:distance_UPA}
    \begin{aligned}
        & r_{mk}   =  r_{1k} \left( 1 + \frac{i(m)^2d_\H^2 + k(m)^2d_\V^2}{r_{1k}^2} \right. \\
        & \quad \left. - \frac{2[i(m)d_\H\sin(\varphi_k)\cos(\theta_k) + k(m)d_\V\sin(\theta_k)]}{r_{1k}} \right)^{\frac{1}{2}} \\ 
        & \quad \overset{(a)}{\approx} r_{1k} + \frac{i(m)^2d_\H^2 + k(m)^2d_\V^2}{2r_{1k}} \\ 
        & \qquad -  i(m)d_\H\sin(\varphi_k)\cos(\theta_k) - k(m)d_\V\sin(\varphi_k)
    \end{aligned}
\end{equation}

\begin{equation}
\label{eq:arrayresponse_UPA}
    \begin{aligned}
        \vect{a}(\varphi,\theta) = [1,\ldots, e^{j\frac{2\pi}{\lambda}[i(n)d_\H\sin(\varphi)\cos(\theta) + k(n)d_\V\sin(\theta)]} \\ ,\ldots,
        e^{j\frac{2\pi}{\lambda}[(M_\H-1)d_\H\sin(\varphi)\cos(\theta) + (M_\V-1)d_\V\sin(\theta)]}]^\T.
    \end{aligned}
\end{equation}

Here, $M_\H$ and $M_\V$ denote the number of antenna elements along the horizontal and vertical axes, respectively, with $d_\H$ and $d_\V$ representing the corresponding inter-element spacing.  Moreover, $i(m) = \mathrm{mod}(m-1,M_\H)$ and $k(m) = \lfloor\frac{m-1}{M_\H}\rfloor$ represent the element indices, and (a) follows from Fresnel approximation~\cite{2024ramezaniefficientmodifiedmusicalgorithm}.

%%%%%%%%%%%%%%%%%%%%%%%%%%%%%%%%%%%%%%%%%%%%%%%%%%%%%%%%%%%%%%%%%%%%%%%%%%%%%%%%%%%%%%%%%%%%%%%%%%%%%%%%%%%%%%%%%%%%%%%%%%%%%%%%%%%%
\section{Background on MUSIC Algorithm}
\label{sec:MUSIC}
In this section, we provide a brief overview of the MUSIC algorithm, which is used to localize signal sources by estimating their AoA and distance~\cite{1991NFMUSIC}. It is important to note that distance estimation is only feasible when the source lies within the near-field region of the sensor array.

The MUSIC algorithm utilizes second-order statistics to detect and localize incoming signals. This is accomplished by estimating both the noise and signal subspaces from the correlation matrix of the received signal. The autocorrelation of the received signal can be expressed as $\vect{R} = \mathbb{E} \{\vect{y}\vect{y}^\H \} = \vect{A}\Tilde{\vect{X}}\vect{A}^\H + \sigma^2\vect{I}_M$, where $\Tilde{\vect{X}} = \mathbb{E}\{\vect{Bx}\vect{x}^\H\vect{B}^\H\}$. When the number of users is smaller than the number of antenna elements, i.e., $K < M$, and the source signals are not fully correlated, the correlation matrix can be decomposed into a combination of noise and signal subspaces, as follows:
\begin{equation}
\label{eq:R}
    \vect{R} = \vect{U}_s\Lambda_s\vect{U}_s^\H + \vect{U}_n \Lambda_n \vect{U}_n^\H,
\end{equation}
where $\vect{U}_n \in \mathbb{C}^{M \times (M-K)}$ and $\vect{U}_s \in \mathbb{C}^{M \times K}$ represent the noise and signal subspaces, respectively. The matrices $\vect{\Lambda}_s$ and $\vect{\Lambda}_n$ contain the eigenvalues corresponding to the eigenvectors in $\vect{U}_s$ and $\vect{U}_n$, respectively. 
%Additionally, $\vect{\Lambda}_s = \diag([\beta_1\rho_1, \ldots, \beta_K\rho_K])$ 

Since the noise subspace is orthogonal to the signal subspace, we have $\vect{U}_n^\H \vect{A} = \vect{0}$. This relationship can be rewritten as $\vect{a}(\boldsymbol{\psi_k})^\H \vect{U}_n \vect{U}_n^\H \vect{a}(\boldsymbol{\psi_k}) = 0$ for $k = 1,\ldots,K$.  Therefore, to determine the user locations, the MUSIC spectrum, as defined in \eqref{eq:MUSIC_spectrum}, is evaluated for various values of distance and angles. The denominator in this expression represents the energy of $\vect{a}(\boldsymbol{\psi})$ projected onto the noise subspace. The $K$ peaks in the spectrum correspond to the locations of the $K$ users.

\begin{equation}
\label{eq:MUSIC_spectrum}
    f(\boldsymbol{\psi}) = \frac{1}{\vect{a}(\boldsymbol{\psi})^\H \vect{U}_n \vect{U}_n^\H \vect{a}(\boldsymbol{\psi})}
\end{equation}

However, calculating the MUSIC spectrum by exhaustively searching over all possible angles and distance values is computationally expensive. The complexity is particularly high for distance resolution, as the search space can range from a few centimeters to several hundred meters, depending on the array aperture, and operating frequency. Therefore, the goal of this paper is to propose a new computationally efficient method based on MUSIC that accurately determines user locations while reducing the computational burden.

\section{Low-Complexity Near-Field Localization Using Angular Domain MUSIC with a Single Anchor Point}
\label{sec:proposed_localization_method}
In the near-field region of a ULA, the array response depends on both distance and azimuth angle. This information is sufficient for determining a user’s location in 2D space using a single anchor point with a large aperture. As discussed earlier, the MUSIC algorithm can estimate the user’s location by performing an exhaustive search over both azimuth and distance. Although 2D MUSIC is asymptotically optimal, it entails significant computational complexity~\cite{1991NFMUSIC}. In this section, we introduce a novel approach that reduces computational complexity by eliminating the need for a distance search.

\begin{figure}
    \centering
    \includegraphics[width=0.7\linewidth]{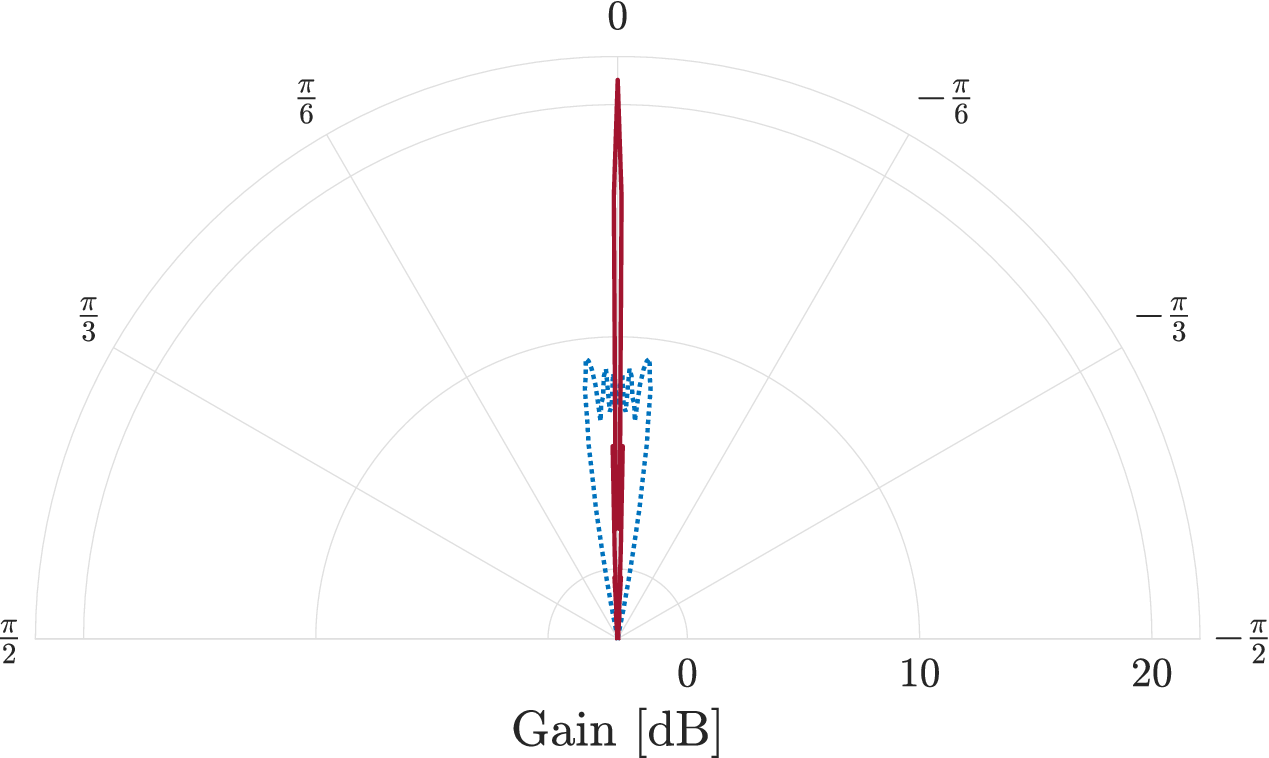}
    \caption{In the far-field of the array, beamforming produces a distinct peak at the angle corresponding to the user’s location. However, in the near-field, the (far-field) beamforming gain is spread across multiple angles, indicating that the angle of arrival varies across the array. This near-field behavior is illustrated by the dotted blue line in the figure.}
    \label{fig:angles_over_aperture}
\end{figure}

When a user is positioned in the near-field region of an antenna array, the AoA varies for each array element, resulting in different AoAs across the entire aperture. This phenomenon is depicted in Figure~\ref{fig:angles_over_aperture} where a transmitter is located at the boresight of a $128$-element ULA with with $\lambda/2$ inter-element spacing. The solid red curve represents the received beamforming gain for a signal source in the far-field, where a narrow beam is shaped at a specific angle, indicating a uniform AoA across all elements, and the array response depends on a single angle. In contrast, for a signal source in the near-field, the array cannot form a single beam with a far-field beamformer in one direction; instead, the beam is spread across multiple angles due to the varying AoAs perceived by different elements of the array\footnote{It is important to note that in Figure~\ref{fig:angles_over_aperture}, a far-field beamformer is used to demonstrate the variation in angles. However, in practical scenarios, the far-field beamformer should be replaced with near-field beamforming (or beamfocusing), which depends on both distance and angle, in order to maximize the gain~\cite{2023RamezaniNear}.}. 

The variation in angles across a single aperture can be leveraged to efficiently localize a signal source in the near-field. To achieve this, the aperture is divided into multiple sub-arrays, effectively transforming the near-field signal source into a far-field source for each sub-array. Since the AoA changes across the aperture, each sub-array detects the source from a different angle. This outcome is akin to having separate access points, each perceiving distinct AoAs. In such cases, the source’s location can be determined through triangulation using the AoAs at each AP. Similarly, by estimating angles from the sub-arrays, triangulation can be applied to localize the user. This approach offers a computational advantage over standard MUSIC~\cite{1991NFMUSIC} or Modified MUSIC~\cite{2012FFNFMUSIC,2024ramezaniefficientmodifiedmusicalgorithm} by focusing solely on angle estimation, rather than conducting a search over both angle and distance.
%the source’s location can be determined by finding the intersection of lines emanating from each access point. Likewise, by estimating angles from the sub-arrays, the Least Squares method can be used to localize the user by solving for the intersection of these lines.

%is akin to having multiple, separately deployed access points, each perceiving distinct AoAs. In the case of multiple access points, the user's location can be determined through triangulation based on the estimated AoAs at each AP. Similarly, when a single aperture observes a near-field user from different angles, it can estimate the user's position through triangulation. 

Let us assume that the antenna aperture is divided into $Q$ sub-arrays, with each sub-array containing $N = M/Q$ elements. Let $\vect{y}_q$ represent the received signal for sub-array $q$, and its corresponding correlation matrix be denoted as $\vect{R}_q = \mathbb{E}\{ \vect{y}_q \vect{y}_q^\H\}$. As in \eqref{eq:R}, this correlation matrix can be expressed as the sum of the signal and noise spaces, provided that $N > K$ and the signal sources are uncorrelated: 
\begin{equation}
    \vect{R}_q = \vect{U}_{sq}\Lambda_{sq}\vect{U}_{sq}^\H + \vect{U}_{nq} \Lambda_{nq} \vect{U}_{nq}^\H.
\end{equation}
In this expression, the subscripts $sq$ and $nq$ refer to the signal and noise spaces for the $q$-th subarray, respectively.

Once the correlation matrix $\vect{R}_q$ is calculated for each subarray, the AOAs can be estimated by identifying $K$ peaks in the MUSIC spectrum as given in \eqref{eq:MUSIC_spectrum_1D}. Here, the array response $\vect{a}_q(\varphi)$ represents the far-field response vector for the ULA subarray consisting of $N$ antennas.
\begin{equation}
\label{eq:MUSIC_spectrum_1D}
    f_q(\varphi) = \frac{1}{\vect{a}_q(\varphi)^\H \vect{U}_{nq} \vect{U}_{nq}^\H \vect{a}_q(\varphi)}.
\end{equation}
After estimating $K$ angles from each subarray, the positions of the $K$ sources can be determined using triangulation. In this process, the location of the reference element for each subarray and the corresponding estimated AoA are known. The signal source positions are then calculated as the intersection points of these lines. The detailed steps for estimating user locations via triangulation are outlined in Algorithm~\ref{alg:Triangulation}. In steps $4-7$, the algorithm identifies $K$ angles and sorts them for each subarray, a critical step to ensure that each angle corresponds to the same source across all subarrays during triangulation. Finally, each source's location is computed as the intersection of lines originating from each subarray center and directed toward the estimated angles. The lines toward source $k$ from each subarray can be represented in vector form as:
\begin{equation}
\label{eq:triangulate}
\left\{\begin{matrix}
 \vect{u}_k  = t_{k1}\cdot \vect{d}_{k1} + \vect{p}_1 \\
 \vect{u}_k  = t_{k2}\cdot \vect{d}_{k2}+ \vect{p}_2 \\
 \vdots\\
 \vect{u}_k  = t_{kQ}\cdot\vect{d}_{kQ} + \vect{p}_Q
\end{matrix}\right.,
\end{equation}
where $\vect{u}_k = [x_k,y_k]^\T$ and $\vect{p}_q = [x_q,y_q]^\T$ denote the location of $k$-th source and $q$-th subarrray in 2D space, and $\vect{d}_{kq} = [\cos(\varphi_{kq}),\sin(\varphi_{kq})]^\T$ represent the direction vector. Moreover, $t_{kq}$ is the scalar parameters to be estimated. By subtracting each row from the first row, we obtain:
\begin{gather}
\label{eq:triangulation2}
\underbrace{\begin{bmatrix}
    \vect{p}_1 - \vect{p}_2\\
    \vect{p}_1 - \vect{p}_3 \\
    \vdots\\
    \vect{p}_1 - \vect{p}_Q
\end{bmatrix}}_{\vect{b} \in \mathbb{R}^{2(Q-1)\times 1}} \!\!=\!\! \underbrace{\begin{bmatrix}
    -\vect{d}_{k1} & \vect{d}_{k2} & 0 & \cdots & 0 \\
    -\vect{d}_{k1} & 0 & \vect{d}_{k3} & \cdots & 0\\
    \vdots & \vdots & \vdots & \vdots & \vdots \\
    -\vect{d}_{k1} & 0 & 0 & \cdots & \vect{d}_{kQ}
\end{bmatrix}}_{\vect{A}_k \in \mathbb{R}^{2(Q-1)\times Q}}\underbrace{\begin{bmatrix}
    t_{k1}\\
    t_{k2} \\
    \vdots \\
    t_{kQ}
\end{bmatrix}}_{\vect{t}_k}\!.
\end{gather}
The estimate for $\vect{t}_k$ can be obtained using Least Squares method:
\begin{equation}
    \hat{\vect{t}}_k = (\vect{A}_k^\T\vect{A}_k)^{-1}\vect{A}_k^\T\vect{b}. 
\end{equation}
Based on \eqref{eq:triangulate}, the source location can be determined relative to each subarray, with the final source position selected as the mean value in step $11$ of Algorithm~\ref{alg:Triangulation}.

\begin{small}
\begin{algorithm}[b!]
\small
\begin{algorithmic}[1]
 \STATE \textbf{Input}: $\vect{y}_q$ for subarray $q = 1,\ldots,Q$ and reference element locations as $\{\vect{p}_1,\ldots,\vect{p}_Q\}$
 \STATE Evaluate $\vect{R}_q = \mathbb{E}\{\vect{y}_q\vect{y}_q^\H\}$ for $q = 1,\ldots,Q$
\STATE {Eigenvalue decomposition of $\vect{R}_q$ and extract $\vect{U}_{nq}$ which are $(N -K)$ Eigenvectors associated with $(N - K)$ lowest Eigenvalues.}
 \FOR{$q = 1,\ldots,Q$}  
    \STATE Evaluate $f_q(\varphi)$ as expressed in \eqref{eq:MUSIC_spectrum_1D}. \\
    \texttt{\footnotesize // Find $K$ peaks and sort the angles}
    \STATE $\{ \varphi_{1q},\varphi_{2q},\ldots, \varphi_{Kq}\} \gets \mathrm{Sort}\left(\mathrm{Peaks}_K f_q(\varphi)\right)$ 
\ENDFOR \\
\texttt{\footnotesize // Triangulation}
\FOR{$k = 1,\dots,K$}
    \vspace{0.1cm}
    \STATE Compute $\vect{A}_k$ and $\vect{b}$ as defined in \eqref{eq:triangulation2}
    \STATE $\hat{\vect{t}}_k = (\vect{A}_k^\T\vect{A}_k)^{-1}\vect{A}_k^\T\vect{b}$
    \STATE $\hat{\vect{u}}_k = 1/Q \left(\sum_{q=1}^{Q} \vect{p}_q + \hat{\vect{t}}_k[q]\cdot \vect{d}_q\right)$  \texttt{\footnotesize // $\hat{\vect{t}}_k[q]: q$-th row}
\ENDFOR
 \STATE \textbf{Output}: Estimated signal source locations $\{\hat{\vect{u}}_1,\hat{\vect{u}}_2,\ldots,\hat{\vect{u}}_K\}$
\caption{Localization using triangulation in 2D space}
\label{alg:Triangulation}
\end{algorithmic}
\end{algorithm}
\end{small}

%%%%%%%%%%%%%%%%%%%%%%%%%%%%%%%%%%%%%%%%%%%%%%%%%%%%%%%%%%%%%%%%%%%%%%%%%%%%%%%%%%%%%%%%%%%%%%%%%%%%%%%%%%%%%%%%%%%%%%%%%%%%

%%%%%%%%%%%%%%%%%%%%%%%%%%%%%%%%%%%%%%%%%%%%%%%%%%%%%%%%%%%%%%%%%%%%%%%%%%%%%%%%%%%%%%%%%%%%%%%%%%%%%%%%%%%%%%%%%%%%%%%%%%%%%%%%%%%%

% 
%%%%%%%%%%%%%%%%%%%%%%%%%%%%%%%%%%%%%%%%%%%%%%%%%%%%%%%%%%%%%%%%%%%%%%%%%%%%%%%%%%%%%%%%%%%%%%%%%%%%%%%%%%%%%%%%%%%%%%%%%%%%
\section{Numerical Results}
\label{sec:numericalRes}
In this section, we present numerical results demonstrating that the proposed method can successfully localize a user when positioned in the near-field region of an antenna array. Additionally, we provide a comparison with standard and Modified MUSIC algorithms~\cite{1991NFMUSIC,2012FFNFMUSIC,2024ramezaniefficientmodifiedmusicalgorithm}.

\subsection{Simulation Setup}
We consider a ULA sensor array equipped with $255$ antenna elements, spaced at $\lambda/4$ intervals. The signal sources are randomly distributed within the near-field region, with distances sampled from $r \sim \mathcal{U}[d_{\mathrm{bjo}},d_{\mathrm{FA}}/10]$, and angles from $\varphi \sim \mathcal{U}[-\pi/3,\pi/3]$. When $\lambda = 0.01\,$m, the minimum and maximum distances are $d_{\mathrm{bjo}} = 1.36\,$m and $d_{\mathrm{FA}}/10 = 8.7\,$m, respectively. Bj{\"o}rnson distance $d_{\mathrm{bjo}} = 2\cdot D$ is defined as the distance beyond which the channel gain remains constant for all antenna elements. The Fraunhofer array distance can be calculated as $d_{\mathrm{FA}} = 2D^2/\lambda$ where, $D$ represents the array maximum length~\cite{2023RamezaniNear}. Additionally, the array is virtually divided into three sub-arrays, and the estimated angles from each sub-array are used for triangulation to determine the source's location.
\subsection{Localization Accuracy in 2D}
\begin{figure}
    \centering
    \includegraphics[width=0.7\linewidth]{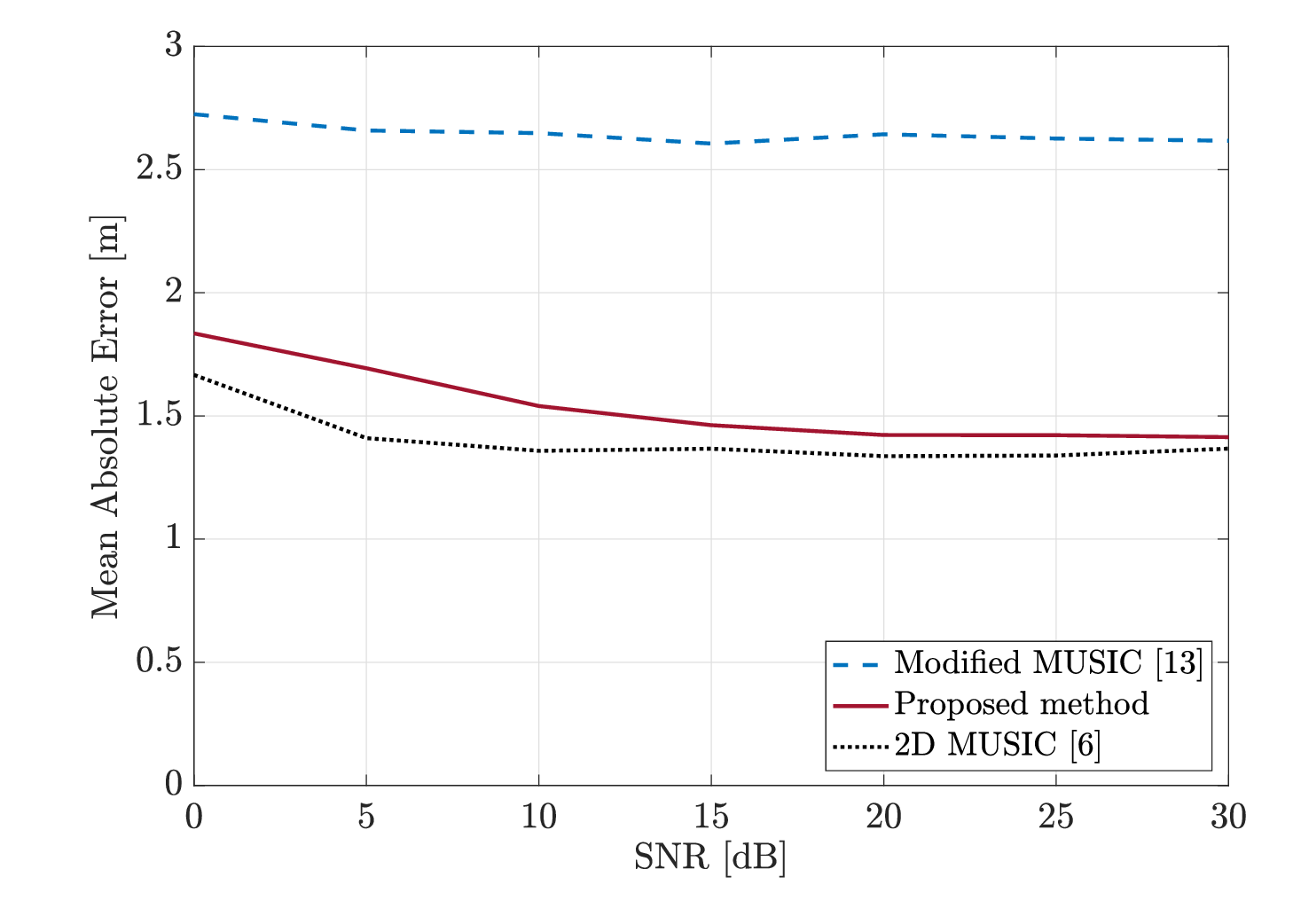}
    \vspace{-0.6cm}
    \caption{The localization error with respect to SNR. For high SNR values, the proposed method reaches the same performance as 2D MUSIC with much less computation involved.}
    \label{fig:error_snr}
\end{figure}

We assess the localization accuracy using the Mean Absolute Error (MAE), which is defined as:
\begin{equation}
    \mathrm{MAE} = \mathbb{E}\{ \| \vect{u} - \hat{\vect{u}} \| \}
\end{equation}
where $\hat{\vect{u}}$ represents the estimated location, and $\vect{u}$ represents the true location of the sources. The results are shown in Figure~\ref{fig:error_snr} for various SNR values in the presence of $K = 6$ concurrent signal sources. The three curves correspond to different algorithms used to localize the user in the near-field region of the array. The baseline algorithm is 2D-MUSIC~\cite{1991NFMUSIC}, an extension of the standard MUSIC algorithm that performs an exhaustive search over both the azimuth and distance planes. Additionally, we implement Modified MUSIC~\cite{2012FFNFMUSIC}, which replaces the 2D search with multiple 1D searches. This approach reduces computational complexity but at the cost of some performance degradation in terms of localization accuracy. The final curve represents the proposed method, detailed in Section~\ref{sec:proposed_localization_method}. 

The search grid for the modified and 2D MUSIC algorithms is limited to $2.9\,$cm in the distance domain and set to $1$ degree in azimuth. We observe that higher SNR values lead to greater accuracy in localizing the signal source. Additionally, the proposed method, despite its lower computational complexity, achieves performance similar to that of 2D MUSIC, particularly in high SNR scenarios, with an error difference of approximately $4\,$cm. In contrast, Modified MUSIC experiences significant performance degradation. This is because this method approximates the distance expression in~\eqref{eq:distance_ULA} to decouple angle from distance, estimating the angle first, followed by distance estimation in a subsequent step.

\begin{figure}
    \centering
    \includegraphics[width=0.7\linewidth]{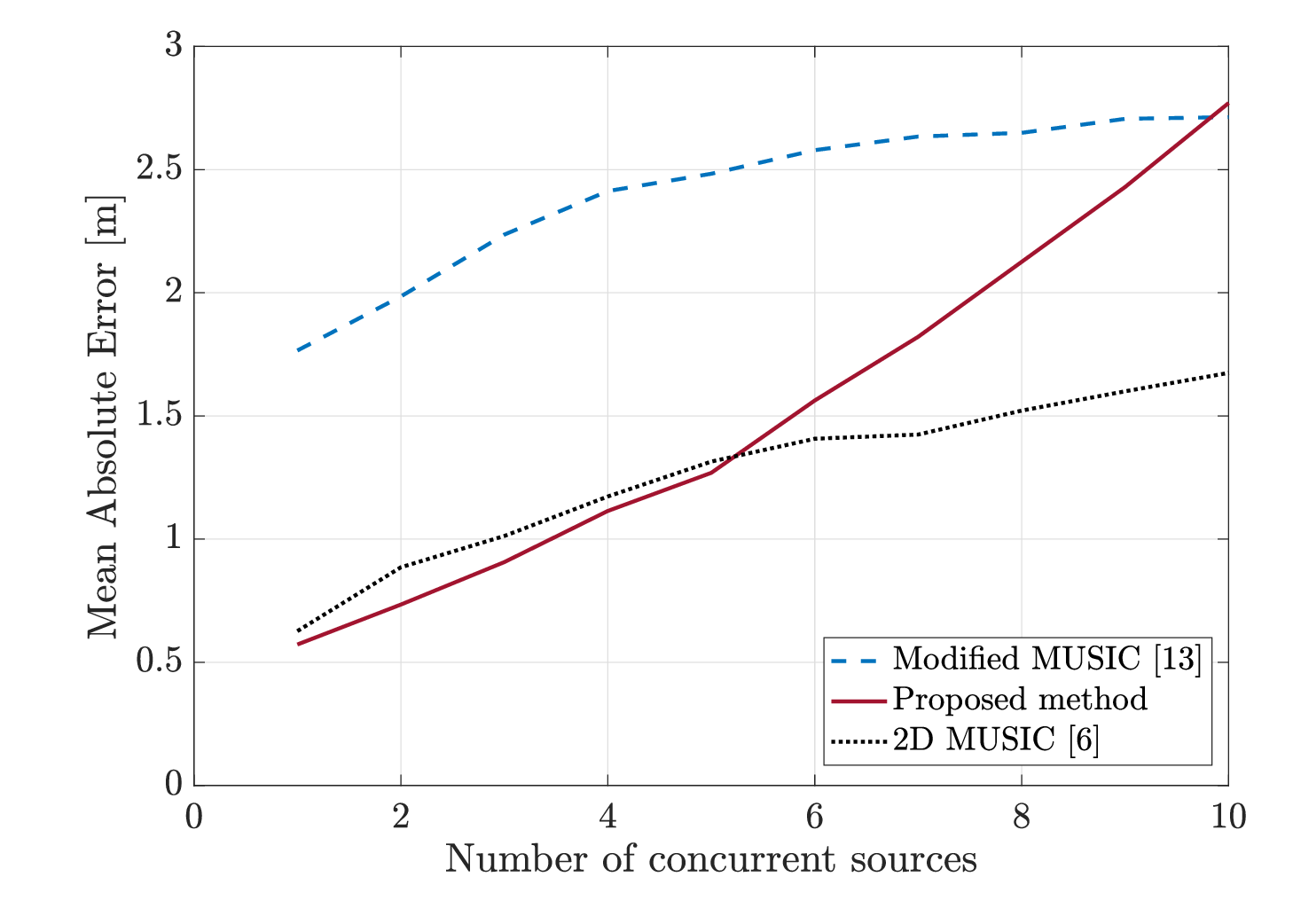}
    \vspace{-0.58cm}
    \caption{Localization error in 2D space with respect to the number of concurrent sources, with SNR fixed at $10\,$dB.}
    \label{fig:error_users}
\end{figure}

Moreover, Figure~\ref{fig:error_users} shows the achievable error for different numbers of concurrent sources with the SNR fixed at $10\,$dB and $15$ snapshots collected to evaluate the correlation matrix. The proposed method outperforms the 2D MUSIC when the number of concurrent sources is fewer than $6$. This advantage stems from the computational limitations of 2D MUSIC in conducting a high-resolution search across both axes. However, as the number of users increases, the performance of the proposed method declines. This degradation can be attributed to two factors. First, the proposed method processes the signal along a single axis, specifically the azimuth, while both Modified and 2D MUSIC resolve incoming signals across the azimuth-distance plane, leveraging both degrees of freedom. Second, the proposed algorithm divides the entire aperture array into sub-arrays, each with a shorter aperture length, resulting in reduced resolution for resolving angles of arrival.

\vspace{-0.1cm}
\subsection{Computational complexity analysis}
Let us consider fixing the search resolution by sampling both the angular and distance domains $M$ times. Applying the current search resolution to our simulation results in the angular grid samples separated by $2\pi/(3M)$, while the distance grid points are spaced by $(d_{\mathrm{FA}}/10 - d_{\mathrm{bjo}}) / M$. Given this resolution, the computational complexity of obtaining the standard MUSIC spectrum in \eqref{eq:MUSIC_spectrum} is $O(2M^3(M-K))$. In contrast, the complexity of evaluating the MUSIC spectrum in \eqref{eq:MUSIC_spectrum_1D} $Q$ times is $O(2M^2(M/Q-K+1))$, which is $M \cdot Q$ times lower than the previous case, where $Q$ represents the number of subarrays into which the aperture is divided.

Furthermore, computing the array response in the near-field is more computationally intensive than in the far-field. The problem gets worse in the 2D search space, where identifying peaks is more complex, while in the 1D search space, it becomes much simpler. Additionally, the Modified MUSIC algorithm has a complexity of $O(2M^2(M-2Q+K))$ for evaluating MUSIC spectrum to estimate the angles~\cite[Eq (23)]{2012FFNFMUSIC}, followed by $O(2M^2(5MK-5K^2))$ for determining all the distances~\cite[Eq (28)]{2012FFNFMUSIC}. In this method, the correlation matrix must be computed twice, further increasing the overall computational load.

To provide a clearer understanding of the differences in computational complexity, Table~\ref{tab:time_consumption} reports the time taken to compute the MUSIC spectrum for each method. As shown, evaluating the 1D MUSIC spectrum $Q$ times takes around $0.1\,$ms, while evaluating the 2D MUSIC spectrum over the azimuth-distance plane takes more than $37\,$ms, highlighting the significant reduction in complexity achieved by our algorithm. Similarly, the proposed method is more efficient than the modified MUSIC algorithm, where the latter takes $1.6\,$ms to evaluate the MUSIC spectrum for both angle and distance estimation. Additionally, the proposed method uses less memory for computation and storage, making it more practical for real-world applications.

\begin{table}[]
\caption{Elapsed time to evaluate the MUSIC spectrum}
    \centering
    \begin{tabular}{c|c|c}
         & Elapsed Time & Time Standard Deviation  \\
         \hline
         \hline
         $2$D MUSIC & $37.4\,$ms & $8.8\,$ms \\
         \hline
         Modified MUSIC & $1.6\,$ms & $1.7\,$ms \\ 
         \hline
         Proposed MUSIC & $0.1\,$ms & $0.01\,$ms \\
    \end{tabular}
    
    \label{tab:time_consumption}
\end{table}
\vspace{-0.1cm}
\subsection{Localization accuracy in 3D}
Algorithm~\ref{alg:Triangulation} can be easily adapted to localize users in 3D space using a UPA. The methodology remains consistent: a large UPA can be divided into $Q$ UPA-structured subarrays, each estimating both azimuth and elevation angles. With these angle estimates and the positions of each subarray, user locations can be determined via triangulation. 

\begin{figure}
    \centering
    \begin{overpic}[width=0.7\linewidth]{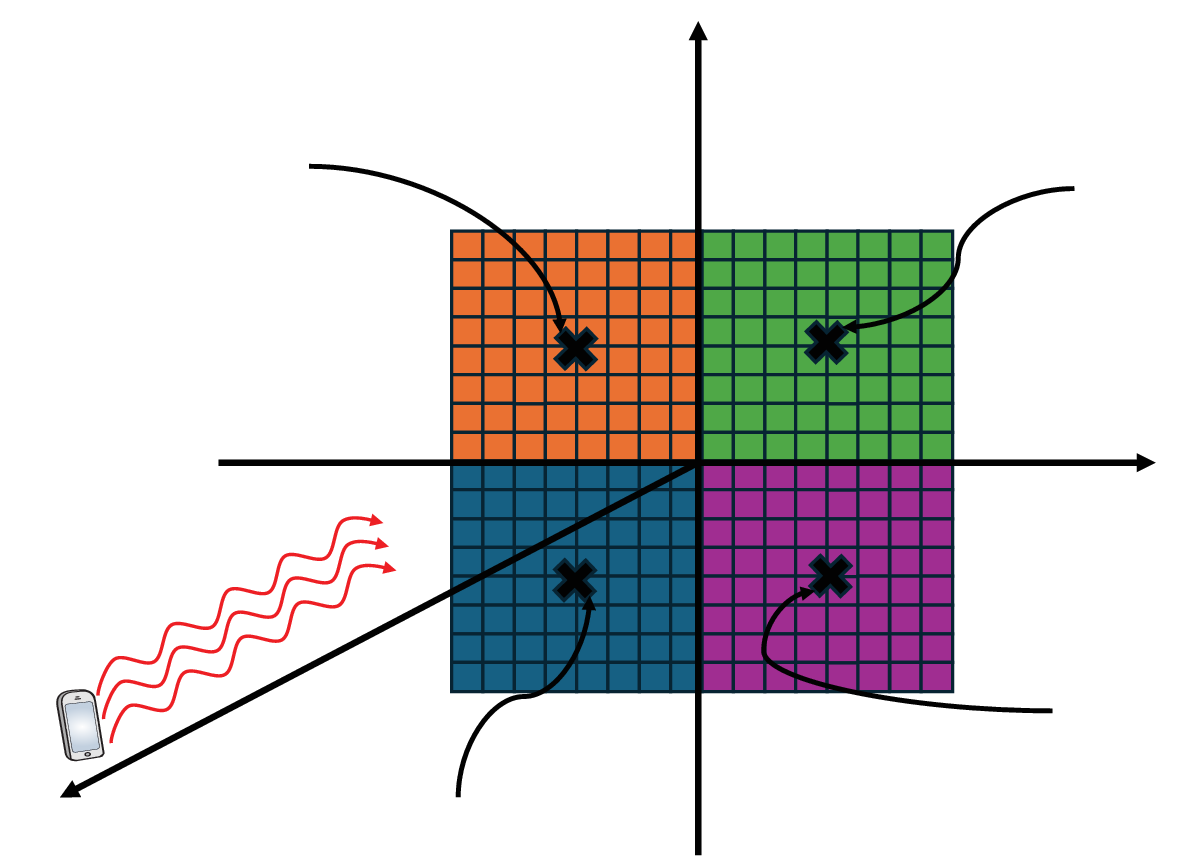}
        \put(26,2){$(x_1,y_1)$}
        \put(80,7){$(x_2,y_2)$}
        \put(15,61){$(x_3,y_3)$}
        \put(80,60){$(x_4,y_4)$}
        \put(0,24){Signal}
        \put(0,20){source}
    \end{overpic}
    \vspace{-0.1cm}
    \caption{A $16 \times 16$ UPA is divided into four sub-arrays, each consisting of $8 \times 8$ elements. Although the signal source is positioned in the near-field of the entire UPA aperture, it lies in the far-field relative to each sub-array.}
    \label{fig:UPA_split}
\end{figure}
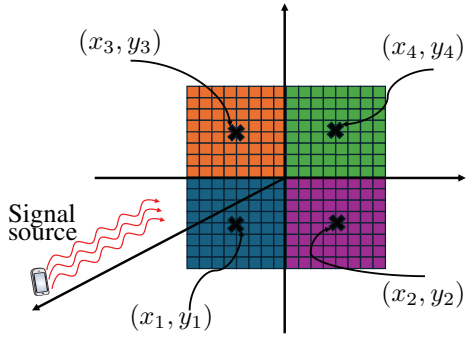

We now present numerical results for a UPA-based sensor array. In this case, the UPA consists of $16\times 16$ antennas with an inter-element spacing of $\lambda/4$. Assuming $\lambda = 0.3\,$m, the resulting Fresnel region lies between $0.07\,$m and $1.9\,$m. We consider two concurrent sources within this region, where the spherical wavefront is captured by the aperture and used for localization. The proposed method divides the entire aperture into four subarrays, each containing $8 \times 8$ elements, as shown in  Figure~\ref{fig:UPA_split}. Each subarray estimates the angles (azimuth and elevation) using MUSIC, performing a grid search over both planes. These estimated angles are then used to localize the user through triangulation, with the center of each sub-array serving as the reference point.

\begin{figure}
    \centering
    \includegraphics[trim={1.2cm 0 2cm 0},clip, height=4.5cm,width=7cm]{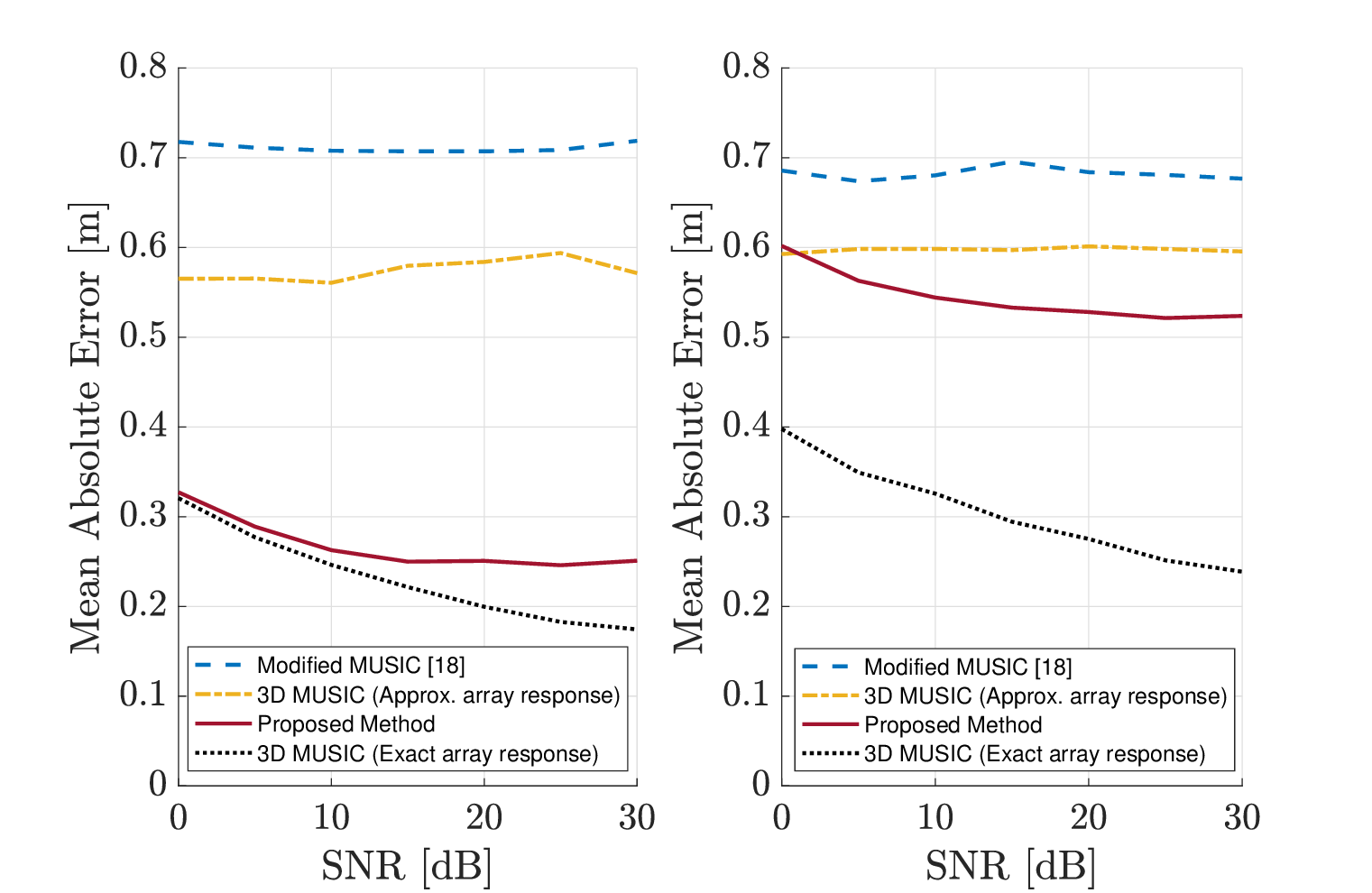}
    \vspace{-0.3cm}
    \caption{3D localization error as a function of SNR for a UPA sensor array: the left figure illustrates results with two concurrent sources, while the right figure shows results with four concurrent sources.}
    \label{fig:SNR3D}
\end{figure}

Figure~\ref{fig:SNR3D} shows the localization accuracy at various SNR levels for two scenarios: with two concurrent signal sources on the left figure and four on the right. The 3D MUSIC algorithm conducts a full 3D search over azimuth, elevation, and distance, which is computationally intensive. The figure shows two curves for 3D MUSIC—one using the approximation in~\eqref{eq:distance_UPA} to simplify array response calculations, and the other using the exact expression to calculate distance, though this increases computational load. Modified MUSIC also uses the same approximation from~\eqref{eq:distance_UPA} and relies on the anti-diagonal elements of the correlation matrix to estimate angles and distance separately~\cite{2012FFNFMUSIC,2024ramezaniefficientmodifiedmusicalgorithm}. In the figure, the proposed algorithm outperforms both Modified MUSIC and 3D MUSIC with the approximated array response, all while significantly reducing computational complexity. Furthermore, it achieves performance comparable to 3D MUSIC with the exact array response, particularly when two concurrent sources are present. The reduced performance observed with four concurrent sources follows the same trend shown in Figure~\ref{fig:error_users}, indicating that as the number of concurrent sources increases, performance declines.

\begin{comment}
  \begin{figure}
    \centering
    \includegraphics[width = \linewidth]{Figures/SNR3DCorrected.eps}
    \vspace{-0.8cm}
    \caption{Localization error in 3D space as a function of SNR for a UPA sensor array.}
    \label{fig:SNR3D}
\end{figure}

\begin{figure}
    \centering
    \includegraphics[width=\linewidth]{Figures/SNR3D.eps}
    \caption{Caption}
    \label{fig:enter-label}
\end{figure}  
\end{comment}

%Now, we extend our preliminary's results into $3\mathrm{D}$ space where an antenna array is generally a function of distance and two angles, e.g., azimuth and elevation. In this more complex scenario, the proposed method run a $2\mathrm{D}$ search grid over azimuth-elevation space while the exhaustive search runs a $3\mathrm{D}$ search which is computationally very expensive and prohibitive and not scalable.

%In this scenario we consider a $128 \times 128$ UPA with inter-element spacing of $\lambda/2$ over horizontal and vertical axes. In addition, the users are randomly located in front of the antenna where $\varphi \sim [-\pi/3,\pi/3]$, $\vartheta \sim [-\pi/3,\pi/3]$ and $r \sim \mathcal{U}[d_{\mathrm{bjo}},d_{\mathrm{FA}}]$. In this setup $d_{\mathrm{bjo}} = $ and $d_{\mathrm{FA}} = $.

%Figure illustrates the localization accuracy based on one anchor point when the user has fallen within the near-field range.

%%%%%%%%%%%%%%%%%%%%%%%%%%%%%%%%%%%%%%%%%%%%%%%%%%%%%%%%%%%%%%%%%%%%%%%%%%%%%%%%%%%%%%%%%%%%%%%%%%%%%%%%%%%%%%%%%%%%%%%%%%%%%%%%%%%%
\vspace{-0.1cm}
\section{Conclusions}
\label{sec:conclusion}
\vspace{-0.1cm}
In this paper, we present a novel method based on the MUSIC algorithm for localizing sources in the near-field region of a sensor array, eliminating the need for a grid search over distance. This is accomplished by leveraging the variation in angular patterns across the aperture, allowing localization solely through angle-of-arrival estimation at different sub-arrays and triangulation. Our method surpasses Modified MUSIC in both accuracy and computational efficiency, while delivering accuracy comparable to conventional 2D-MUSIC and 3D-MUSIC when the array is structured as a ULA or UPA, respectively.

\vspace{-3mm}
\bibliographystyle{IEEEtran}
\bibliography{IEEEabrv,refs}

\end{document}